\RequirePackage{fix-cm}
\documentclass[smallextended]{svjour3}       % onecolumn (second format)
\smartqed  % flush right qed marks, e.g. at end of proof
\usepackage{amssymb}
\usepackage{amsmath}
\usepackage{graphicx}
\usepackage{color}
\usepackage{amssymb}

\usepackage{graphicx}
\usepackage{amsmath}
\usepackage{epsfig}
\usepackage{color}
\usepackage[utf8]{inputenc}
\usepackage[T1]{fontenc}
\usepackage{array,multirow,makecell}
\setcellgapes{1pt}
\newcolumntype{R}[1]{>{\raggedleft\arraybackslash }b{#1}}
\newcolumntype{L}[1]{>{\raggedright\arraybackslash }b{#1}}
\newcolumntype{C}[1]{>{\centering\arraybackslash }b{#1}}

\newtheorem{axiom}[theorem]{Axiom}

\newtheorem{principle}[theorem]{Principle}

\journalname{Ricerche di Matematica}

\begin{document}

\title{Principle of virtual action  in continuum mechanics}

\titlerunning{Principle of virtual action}        % if too long for running h
\author{\author{Henri Gouin}}

\institute{ \at
           Aix Marseille University, CNRS, IUSTI, UMR 7343, Marseille, France.\newline  E-mails: henri.gouin@ens-lyon.org; henri.gouin@univ-amu.fr} 


\maketitle

\begin{abstract}
We present the principle of virtual action as  a foundation of continuum mechanics.  Used mainly in relativity, the method has a useful application in classical mechanics and places the notion of action as the basic concept of dynamics. The principle is an extension of virtual work to space-time. It extends the efforts made by d'Alembert and Lagrange.   Unlike the classical case of equilibrium, the principle of virtual action becomes a  postulate for the formulation of models in dynamics.  It allows to use a minimal set of clear conjectures and is extended to the case of media with dissipation; it can be used for more complex systems.
	\subclass 
{75A05 \and 75-10
  \and 76-10}
\end{abstract}
\journalname{Ricerche di Matematica}

 \section{Introduction}
For  continuous media, the principle of virtual work is   convenient to obtain   the equations of equilibrium   \cite{Serrin,Casal,Marsden,Gouin1,Germain}.  Forces  and more generally  stresses are  dual   of   displacements and the principle yields  equilibrium equations.
It can also be used to obtain the equations of conservative motions: it is enough to add  inertial forces to the different efforts.\\
 However, since motions are associated with space--time quantities, it is  natural to use the notion of action rather than  the notion of energy \cite{Gouin3,Gavrilyuk1,Gavrilyuk2}. Indeed, all quantities involved in dynamics such as momentum and inertial forces are derived from the action. 
A principle of virtual action is more efficient than the principle of virtual work because it directly gives not only the equation of momentum  but also the  equation of energy. The thermodynamics being introduced with the greatest simplicity, we naturally obtain the equation of entropy.
 This way of thinking, not   usual in classical mechanics, is the basis  of general relativity which uses  same concepts as classical mechanics but differs in the choice of dynamic behavior laws \cite{Souriau}.  \\
 
 \noindent The motion of a continuous medium is governed by an application
 $\varphi_t$ from $\mathcal{D}_0$ into $\mathcal{D}_t$, where $\mathcal{D}_0$ is an open set of a three dimensional domain  and $\mathcal{D}_t$ is at time $t$ an open set of the three--dimension physical space $\mathcal E$ occupied by the medium  \cite{Serrin};   $\boldsymbol X =[X^1, X^2, X^3]^T$ denotes the reference position in Lagrange variables, $\boldsymbol x =[x^1, x^2, x^3]^T$ denotes the particle position in Euler variables at time $t$, where $^T$ denotes the transposition; we write 
 $$\boldsymbol x=\varphi_t(\boldsymbol X)$$
We can also write the motion in the form

 \begin{equation}
 	\left\{
 	\begin{array}{l}
 		\displaystyle 
 		t  = t(\lambda,\boldsymbol X)\\ \qquad\qquad\qquad\qquad{\rm or } \qquad\qquad\boldsymbol z = \mathcal{F(\boldsymbol Z )} \\
 		\displaystyle\boldsymbol x=\boldsymbol x(\lambda,\boldsymbol X)    
 	\end{array}\right.\label{Motion}
 \end{equation}
 where $\lambda$ is a scalar parameter and 
 \begin{equation*}
 	 \boldsymbol z=\left[ \begin{matrix}
 	t\\
 	\boldsymbol x 
 	\end{matrix}\right],\quad\boldsymbol Z=\left[ \begin{matrix}
 		\lambda\\
 		\boldsymbol X 
 	\end{matrix}\right],
 \end{equation*}
 with $\,\ z^0 =t,\ z^i = x^i,  \,\ Z^0 = \lambda,\ Z^i = X^i,\,\ i\in \{1,2,3\} $.\\

 \noindent
 Between times $t_1$ and $t_2$,  the current domain $\mathcal D_t$ of element $\displaystyle \left[ 
 t,
 \boldsymbol x 
 \right]^T$ generates  a manifold $\mathcal W$ of the 4-dimension affine space--time occupied by the medium and $\mathcal W_0 =\mathbb R \times \mathcal D_0 $ becomes the parameter space.
 The motion $\mathcal F$ of the continuous medium,  corresponding to a diffeomorphism on   manifold with edge, is  defined on an open set  so that $\mathcal W$ and $\mathcal W_0$ have  compact closures and piece--wise differentiable boundaries \cite{Gouin1,Gavrilyuk1}.
 We note the similarity with the motion defined by an application between two 4-dimensional spaces and the statics position defined by an application between two 3-dimensional spaces.
 \section{Principle of virtual action}
 \subsection{Virtual motions and virtual displacement of a continuous medium }
 \begin{definition}
 	Consider a differentiable application $\mathcal G$ such that
 	\begin{equation*}
 	(\boldsymbol Z, \varepsilon)\in \mathcal W_0 \times \mathcal O \longrightarrow\boldsymbol  z=\mathcal G(\boldsymbol Z, \varepsilon)\in \mathcal W\quad {\rm with}\quad \mathcal G(\boldsymbol Z, 0) = \mathcal F(\boldsymbol Z) 
 	\end{equation*}
 	where $\mathcal O$ is an open set of $\mathbb R$ containing $0$. We call $\mathcal G$ a one-parameter family of \textit{virtual motions}.\\
 	The associated \textit{virtual displacement} is the vector field 
 	\begin{equation*}
 	\boldsymbol z \in \mathcal W\longrightarrow \boldsymbol\zeta\qquad {\rm where}  \qquad \boldsymbol\zeta =\frac{\partial \mathcal G (\boldsymbol Z, 0)}{\partial \varepsilon},
 	\end{equation*}
which belongs to the vector space, tangent to $\mathcal W$ at point $\boldsymbol z=\mathcal{F(\boldsymbol Z )}$.  In fact we can also write $\boldsymbol\zeta$ as a differential of $\mathcal G(\boldsymbol Z, \varepsilon)$ when $\varepsilon =0$
 	\begin{equation*}
 		\boldsymbol\zeta = d\boldsymbol z \quad {\rm with}  \quad d\boldsymbol Z=\boldsymbol 0\,\ {\rm and}\,\ d\varepsilon = 1
 	\end{equation*}
 \end{definition}
 A linear functional of the virtual displacement is called virtual action and is denoted by $\delta a$. To describe the most classical model, we choose the expression
 \begin{equation*}
 	\delta a =\int_{\mathcal W}\left[S_i \zeta^i+T_i^j\zeta^i_{,j}\right] dw
 \end{equation*}
where $dw$ is the {\it volume element} of ${\mathcal W}$. In matrix notation, we obtain
 \begin{equation}
 	\delta a =\int_{\mathcal W}\left[\boldsymbol S ^T\boldsymbol \zeta+ {\rm Tr}\left(\boldsymbol T \frac{\partial\boldsymbol\zeta }{\partial\boldsymbol z}\right)\right] dw\label{delta a}
 \end{equation}
 which introduces a field of covectors  $\boldsymbol S ^T$ (or forms) and a field of matrices  $\boldsymbol
 T$  which are separated into time and space parts 
 \begin{equation*}
 	\boldsymbol S ^T=\left[-\mathcal P, \boldsymbol F ^T\right]\quad {\rm and}\quad\boldsymbol T = \left[ \begin{matrix}
 		\ \	-\ e\,  ,\ \ \boldsymbol P ^T\\
 		-\boldsymbol U, \ \   \boldsymbol \varphi 
 	\end{matrix}\right]
 \end{equation*}
 where $\mathcal P$, $e$ are scalars, $\boldsymbol F ^T$, $\boldsymbol P ^T$ are linear forms, $\boldsymbol U$ is a vector and  $\boldsymbol \varphi$ is a $3\times 3$ tensor in space $\mathcal{D}_t$. In components,   $i, j\in \{1, 2, 3\}$ 
 \begin{equation*}
 	S_0 = -\mathcal P,\,\ S_i= F_i, \,\ T_0^0 = - e, \,\ T_0^i =- U^i, \,\ T_i^0 = P_i,\,\ T_i^j =\varphi _i^j
 \end{equation*}
 $\boldsymbol S ^T$ and $\boldsymbol T$ will be given by behavior laws as function of the motion. \\
 For the sake of simplicity, we  only consider a  first gradient  model\cite{Germain}. However, a second gradient model based on a functional of the form
 \begin{equation*}
 	\delta a =\int_{\mathcal W}\left[ S_i  \zeta^i+ T_i^j  \zeta_{,j}^i+ C_i^{jk} \zeta_{,jk}^i  \right] dw
 \end{equation*}
 might be easy to develop as it has been done in statics and would introduce not only new types of forces but also non-usual kinematic quantities \cite{Gouin1}.
 
 \subsection{Physical dimensions of tensor quantities}
Manifolds $\mathcal W_0$ and $\mathcal W$ being contained in no--metric affine spaces, we  specify the physical dimensions of the previously defined quantities 
 \vskip 0.5cm
 \begin{center}
 	\begin{tabular}{|R{1.5cm}|C{3.5cm}|L{4.7cm}|}
 		\hline Quantity & Dimension & \qquad\quad Physical nature \\
 		\hline 
 		\hline $\delta a\quad\ $ & $ML^2T^{-1}$ &  Action \\
 		\hline  $dw\quad\, $  & $L^3T$ & Volume element of space--time \\
 		\hline 
 		$\mathcal P\quad\ $  & $ML^{2}T^{-3}.L^{-3}$ & Power per volume \\
 		\hline 
 			${\boldsymbol F}^T\quad\ $  & $ML T^{-2}. L^{-3}$ & Volume force \\
 		\hline 
 		$e\quad\,\ $  & $ML^{2}T^{-2}.L^{-3}$ & Volume energy \\
 		\hline
 		$\boldsymbol U\quad\   $  & $ML^2 T^{-2}.LT^{-1}.L^{-3}$ & Volume energy flux \\
 		\hline
 		$\,\ \boldsymbol P^T\quad   $  & $MLT^{-1}.L^{-3}$ & Volume momentum\\
 		\hline
 		$ \boldsymbol \varphi\quad\,\ $  & $MLT^{-2}.L^{-2}$ &\,Stress tensor\\
 		\hline\hline
 		
 	\end{tabular}
 \end{center}
 \vskip 0.5cm
 We give to these quantities the following names \\
 \begin{equation*}
 	\left | \begin{array}{l}
 		\displaystyle {\rm Source}\quad
 		\boldsymbol S ^T    \\
 		\\{\rm Energy\ source}\quad
 		\mathcal P\\
 		\\ {\rm Momentum\ source}\quad
 		{\boldsymbol F}^T \\
 		\\  {\rm  Stress-energy \ tensor}\quad
 		\boldsymbol T    
 	\end{array}\right.\qquad\qquad \left | \begin{array}{l}
 		\displaystyle
 		{\rm Total\ energy}\quad
 		e \\
 		\\
 		{\rm Energy\ flux\ vector} \quad
 		\boldsymbol U\\  
 		\\
 		{\rm Momentum}\quad
 		\boldsymbol P^T \\
 		\\
 	{\rm Total\ stress}\quad
 		\boldsymbol \varphi
 	\end{array}\right.
 \end{equation*}
 \subsection{The principle of virtual action}
 \begin{principle}
 	For any virtual displacement, null on the edge $\mathcal {\partial W}$ of $\mathcal W$, the virtual action is zero.
 \end{principle}
 i.e., \begin{equation*}
 	\forall\,  \boldsymbol z \in \mathcal W\longrightarrow \boldsymbol\zeta\,, \quad {\rm with}  \quad \boldsymbol\zeta = \boldsymbol 0 \quad {\rm  if}  \quad \boldsymbol z \in \mathcal {\partial W}\,   , \quad  {\rm then}  \quad \delta a =0
 \end{equation*}
 This can also be stated in an equivalent way \\
 
 \textit{For any virtual motion with compact support, the virtual action is zero.}
 
 \section{The equations of dynamics }
 \subsection{Equations of energy and motion}
 In this section, we consider $\boldsymbol T$ as a differentiable tensor. By integration by parts we obtain 
 \begin{equation}
 	\delta a =\int_{\mathcal W}\left[\boldsymbol S ^T-   {\rm Div}\left(\boldsymbol T\right)  \right] \boldsymbol \zeta\,dw+\int_{\partial\mathcal W}\boldsymbol N^T\boldsymbol T\,\boldsymbol \zeta\, d\tau\label{IPP}
 \end{equation}
 or
 \begin{equation*}
 	\delta a =\int_{\mathcal W}\left[S_i-   T^j_{i,j} \right]  \zeta^i\,dw+\int_{\partial\mathcal W}  N_j T_i ^j\,  \zeta^i\, d\tau \qquad \{i,j = 0,1,2,3\}
 \end{equation*}
 Covector $\boldsymbol N^T \,\ (N_j \,   \{j = 0,1,2,3\}) 
 $ 
 is the annulator  of vectors belonging to the hyperplane tangent to   $\partial \mathcal W$, edge of $\mathcal W$ with \textit{surface element} $d\tau$.\\
 The fundamental lemma of  calculus of variations associated with the principle of virtual action gives the equation 
 \begin{equation}
 	{\rm Div}\,\boldsymbol T =\boldsymbol S ^T\qquad {\rm or}\qquad T^j_{i,j}= S_i\qquad \{i,j = 0,1,2,3\}\,, \label{delta a1}
 \end{equation}
where\,  Div\,  denotes the divergence operator in space--time $\mathcal W$.  
 This equation can be separated in terms of time and space \\
 The term in time 
 \begin{equation*}
 	\frac{\partial e}{\partial t}+ {\rm div}\, \boldsymbol U =\mathcal P\qquad {\rm or}\qquad \frac{\partial e}{\partial t}+ U_{,i}^i  =\mathcal P\qquad \{i = 1,2,3\}\,,
 \end{equation*}
 where\,  div\,  denotes the divergence operator in space $\mathcal{D}_t$,
 is the equation of energy.  The energy of a continuous medium is described by a quadri--vector $\displaystyle \boldsymbol E = \left[ \begin{matrix}
 	e\\
 	\boldsymbol U 
 \end{matrix}\right]$, whose divergence in space--time is the source of energy\,\ ${\rm Div}\, \boldsymbol E= \mathcal P$    (energy introduced per unit of time into  unit of volume).\\
 The terme in space  is 
 \begin{equation*}
 	\frac{\partial \boldsymbol P^T}{\partial t}+ {\rm div}\,   \boldsymbol \varphi = \boldsymbol F^T\qquad {\rm or}\qquad 	\frac{\partial P_i}{\partial t}+\varphi_{i,j}^j = F_i\qquad \{i = 1,2,3\}\label{geneq}
 \end{equation*}
 which is the momentum equation. The momentum is described by a $(3\times4)$--tensor $\displaystyle {\Large\boldsymbol\pi} = \left[ \begin{matrix}
 	\boldsymbol P^T\\
 	\boldsymbol \varphi 
 \end{matrix}\right]$ whose divergence in space--time \,\	${\rm Div}\,   {\Large\boldsymbol\pi}= \boldsymbol F^T$\,\  is the source of momentum (momentum communicated per unit of time into  unit of volume). Equations \eqref{delta a1} are conservation laws in continuum mechanics \cite{Lamb,Dafermos,Capobianco}.
 \subsection{Relation between equation  of energy and equation of motion}
Application $\mathcal F$ in Eq. \eqref{Motion} defines a $\lambda$-parameterized motion. If   its variation  only consists in a change of  parameter $\lambda$, the motion remains unchanged and the corresponding virtual action is null. For such an application  $\boldsymbol \zeta$ is proportional to space--time velocity  $\mathcal V = \left[ \begin{matrix}
 	1\\
 	\boldsymbol v
 \end{matrix}\right]$ (where $\boldsymbol v$ is the spatial velocity) and assumed null on ${\mathcal\partial W}$. The   functional \eqref{delta a} is identically zero and   
 \begin{equation}
 	\left[\boldsymbol S ^T-   {\rm Div}\,\boldsymbol T   \right] \mathcal V \equiv 0\label{eq entropy}
 \end{equation}
  is an identity which can be written as 
 \begin{equation*}
 	\frac{\partial e}{\partial t}+ {\rm div}\, \boldsymbol U -\mathcal P-\left[\frac{\partial \boldsymbol P^T}{\partial t}+ {\rm div}\,   \boldsymbol \varphi -\boldsymbol F^T\right]\boldsymbol v\equiv 0
 \end{equation*}
or
 \begin{equation*}
 	\frac{\partial e}{\partial t}+   U_{,i}^i -\mathcal P-\left[\frac{\partial \boldsymbol P_i}{\partial t}+    \varphi_{i,j} ^j-\boldsymbol F_i\right]  v^i\equiv 0
 \end{equation*}
 \begin{theorem}
 	The equation of energy is obtained by multiplying the  equation of momentum by the velocity $\boldsymbol v$.
 \end{theorem}
In Section 6, we will see that Eq. \eqref{eq entropy} represents the equation of entropy.
 \section{Conservative media}
 For perfect fluids,  hyperelastic media in small or large deformation, and more generally for conservative media   because the energy is not dissipated by viscosity or thermal conduction, the virtual action can be constructed  as follows
 
 \begin{axiom}
 	To any motion  $\boldsymbol Z \longrightarrow \boldsymbol z =\mathcal F(\boldsymbol Z)$  corresponds an action\,\,\,\ $a$\,\,\ localized in space--time $\mathcal W$ by a density\,\ $\mathcal L$\,\ per unit volume and time, defined at any point of $\mathcal W$ and written in integral form, i.e.,
 	\begin{equation*}
 		a =\int_{\mathcal W}\mathcal{L}\, dw
 	\end{equation*}
 	The virtual action $\delta a$ is the variation   of  action $a$.
 \end{axiom}
 \begin{axiom}
 The value of  $\mathcal{L}$ at point $\boldsymbol z$ is a given function of $\boldsymbol z$, $\displaystyle\mathcal B =\frac{\partial\boldsymbol z}{\partial\boldsymbol Z}$ and $\boldsymbol Z$, 
 	\begin{equation}
 		\mathcal{L}= L(\boldsymbol z, \mathcal B, \boldsymbol Z)\label{lagrangien}
 	\end{equation}
 \end{axiom}

\noindent The invariance of $\mathcal{L}$ with respect to the parameterization   implies that $\mathcal{L}$ depends on $ \boldsymbol Z$  only through $ \boldsymbol X$  and on $\mathcal B$ only through $\mathcal A$, where $\displaystyle \mathcal A = \left[ \begin{matrix}
 \ 	1  ,\     \boldsymbol 0 ^T\\
 	\boldsymbol v\, , \,    \boldsymbol F 
 \end{matrix}\right]$, i.e,
 \begin{equation}
 	\mathcal{L}= L_0(\boldsymbol z, \mathcal A, \boldsymbol Z)\equiv L_1(t, \boldsymbol x, \boldsymbol v, \boldsymbol F, \boldsymbol X)\label{Lagrangien1}
 \end{equation} \\
 Thanks to axioms, $\mathcal L$ is a Lagrangian and  the equations of motion \eqref{delta a1} are  equations of Lagrange.\\
 The variation of action\,\ $a$\,\ yields  a source $\boldsymbol S^T$ and a  stress--energy  tensor $\boldsymbol T$. By taking  $\mathcal L$ in the form \eqref{lagrangien}, we obtain 
 \begin{equation*}
 	\boldsymbol{S}^T = \frac{\partial L}{\partial \boldsymbol{z}},\,\ \boldsymbol T= L+\mathcal B\,\frac{\partial L}{\partial\mathcal B}\quad {\rm or}\,\ S_i = \frac{\partial L}{\partial  {z_i}},\,\ \boldsymbol T_i^j= L\,\delta_i^j+ \mathcal B_k^j \frac{\partial L}{\partial\mathcal B_k^i}, \  \{i,j = 0,1,2,3\} 
 \end{equation*}
where $\delta^j_i$ is the Kronecker symbol. If $\mathcal L$ is restricted to  form \eqref{Lagrangien1}, we have  
 \begin{equation}
 	\boldsymbol{S}^T = \frac{\partial L}{\partial \boldsymbol{z}},\quad \boldsymbol T= L+\mathcal A\,\frac{\partial L}{\partial\mathcal A} \Pi\label{formula}
 \end{equation}
where $\Pi$ is the projector parallel to $\mathcal V$ of space--time $\mathcal W$ into space $\mathcal D_t$.
 \begin{equation*}
 	\Pi = \left[ \begin{matrix}
 		\quad\,  0\, ,\     \boldsymbol 0 ^T\\
 		-\boldsymbol v\, , \,    \boldsymbol 1 
 	\end{matrix}\right]\quad{\rm where}\quad \Pi_0^0 = 0,\ \ \Pi_0^j = -v^j,\ \ \Pi_i^0 = 0,\ \ \Pi^j_i =  \delta^j_i ,
 \end{equation*}
 The formulas \eqref{formula} are explained in terms of time and space. If $\mathcal L$ is  in  form $L_1$ of Eq. \eqref{Lagrangien1},
 \begin{equation}
 	\boldsymbol{S}^T =[-\mathcal P, \boldsymbol F^T], \qquad{\rm where}\qquad \mathcal P  =-\frac{\partial L_1}{\partial t},\quad \boldsymbol F^T =-\frac{\partial L_1}{\partial \boldsymbol x}\label{key1}
 \end{equation}
 \begin{equation}
 	\boldsymbol{T}=\left[ \begin{matrix}
 		-e ,\qquad      \boldsymbol P ^T\\ \\
 		\ \ -\boldsymbol U  , \,\     \boldsymbol v  \boldsymbol P ^T- \boldsymbol \sigma
 	\end{matrix}\right]\label{key2}
 \end{equation}
 \begin{equation*}
 	{\rm where}\quad \boldsymbol P^T  = \frac{\partial L_1}{\partial \boldsymbol v},\quad e = \boldsymbol P^T\boldsymbol v -L_1, \quad \boldsymbol{\sigma}= -L_1\boldsymbol 1- \boldsymbol F\frac{\partial L_1}{\partial \boldsymbol F},\quad \boldsymbol U= e\,\boldsymbol v- \boldsymbol{\sigma}\,\boldsymbol v
 \end{equation*}
 In  stress--energy  tensor $\boldsymbol T$, we have replaced    $\boldsymbol \varphi$  by $ \boldsymbol v  \boldsymbol P ^T- \boldsymbol \sigma$. The static stress tensor is $\boldsymbol\sigma$ expressed as a function of $L_1$,  and the Lagrangian $ L_1(t, \boldsymbol x, \boldsymbol 0, \boldsymbol F, \boldsymbol X)$ provides the virtual work expression.
 \begin{theorem}
 	For a conservative motion,  the velocity quadri--vector $\mathcal{V}$ is the eigenvector of the  stress--energy  tensor. The   eigenvalue is the Lagrangian.
 \end{theorem}
Since $\Pi\,\mathcal V = \boldsymbol{0}$, the property comes from the second relation \eqref{formula} and the action is independent of   the parameter $\lambda$. This property of the stress--energy tensor is expressed as energy flux vector
 $ \boldsymbol U$ is  $e \boldsymbol v- \boldsymbol{\sigma}\boldsymbol v$, but is not valid when we have heat conduction.
 \section{The laws of dynamic behavior}
 
 The functional we have chosen as virtual action \eqref{delta a} allows to introduce a number of notions, specifying their tensorial nature.  By \eqref{delta a1} and \eqref{eq entropy}, the principle of virtual action indicates the relations that necessarily link these notions. 
 It is important to note this construction is  only a set of conventions  which obviously cannot  claim to govern the nature. 
The study of   laws of dynamic behavior  can only be carried out when appropriate experiments have been done. In fact, for each concerned medium, the laws express   the different dynamic quantities  we have introduced (in fact $\boldsymbol S^T$ and $\boldsymbol T$) as  function of its motion.
 There is nothing  to restrict the  diversity of these laws, they only must be compatible with relations \eqref{delta a1} and \eqref{eq entropy}.\\
 Note that the virtual work in statics is deduced from the virtual action when the motion is reduced to rest. The dynamic behavior law then provides the static behavior law. \\
 The case of conservative media is particularly simple since it is sufficient to express a behavior law   by giving the Lagrangian. We first examine this case.
 \subsection{Choice of the Lagrangian}
 The Lagrangian summarizes the mechanical properties of the continuous medium and also of  the space--time.

 The Lagrangian depends (see \eqref{Lagrangien1}) 
 
 - On $t$, $\boldsymbol x$ and $\boldsymbol v$  which are quantities defined in space--time,
 
 - On $\boldsymbol X$  which designates the particle position in the reference space $\mathcal D_0$
 
 - On $\boldsymbol F$  which is an application from the tangent vector space $T_{\boldsymbol X}(\mathcal D_0)$ at $\boldsymbol X$   into the tangent vector space $T_{\boldsymbol x}(\mathcal D_t)$ at $\boldsymbol x$.
 \\
 The value of the Lagrangian varies from one particle to another one, or by changing the axis in $T_{\boldsymbol X}(\mathcal D_0)$ and  the Lagrangian $\mathcal L$  depends on ${\boldsymbol X}$ and ${\boldsymbol F}$. If the Lagrangian is independent of ${\boldsymbol X}$, the medium is homogeneous. 
 If the space--time possesses    homogeneity and symmetry,  Lagrangian $\mathcal L$  does not vary with $t$ and ${\boldsymbol x}$ or by changing the axes of $T_{\boldsymbol x}(\mathcal D_t)$ (which has the effect of changing the components of ${\boldsymbol v}$ and ${\boldsymbol F}$).\\
These properties lead to different  models of Lagrangian and experiments conclude  which model is the most appropriate.  The simplest way is to consider the Lagrangian as invariant by displacement in space and by translation in time. To express this property (called the principle of material indifference), we have to introduce a metric in the space. Then, the Lagrangian is  independent of ${\boldsymbol z}$ and depends only  on ${\boldsymbol v}$ by its scalar square ${|\boldsymbol v|^2}$ = ${\boldsymbol v} ^T {\boldsymbol v}$ and on ${\boldsymbol F}$ by the Cauchy tensor ${\boldsymbol G} =   {\boldsymbol F}^T {\boldsymbol F}$, i.e., 
 \begin{equation}
 	{\mathcal L} =L({|\boldsymbol v|^2}, {\boldsymbol G}, {\boldsymbol X})\label{Ind Mat}
 \end{equation}
 In particular, ${\boldsymbol S}^T ={\boldsymbol 0}$ (no source) and  $\boldsymbol\sigma = \boldsymbol\sigma^T$ (the constraint is symmetric; noting that the notion of symmetry is meaningless as long as there is no metric).
 \subsection{Interior and exterior of a medium}
 The study of motions in a terrestrial laboratory shows that the vertical is a privileged direction, which contradicts the previous invariance.
 In general, the influence of material bodies outside the medium destroys the invariance. This influence is expressed by geometrical conditions imposed at the boundary of the medium or by contact actions  whose virtual forms of action are given, and by forces exerted inside the medium. These actions at  distance, of which only two types are currently known - gravitational forces and electro--magnetic forces - are introduced in terms of sources whose virtual actions are added to the  Lagrangian.\\
 It is known that in general relativity, the influence of the exterior is taken into account by the metric of the space--time which is no longer an affine space and the definitions as well as the general equations remain valid when covariant derivatives are used \cite{Souriau}.

 \section{Equations of energy and motion for  continuous media}
 \subsection{Isolated conservative media}
 We consider a Lagrangian   invariant by time translation and  space displacement (material indifference) in form \eqref{Ind Mat}. 
More precisely, ${\mathcal L}$ is the difference between the kinetic and potential volumetric energies 
 \begin{equation}
 	\mathcal L = \rho\left(\frac{1}{2} \,\boldsymbol |\boldsymbol v|^2- \alpha({\boldsymbol G}, s)\right)\label{Lagrangien2}
 \end{equation}
% where $s = s(\boldsymbol X)$ is the specific entropy.
 With such a Lagrangian,  where $\rho$ denotes the density, $\alpha$ the specific energy and $s$ the specific entropy, the principle of virtual action merges with Hamilton's principle \cite{Gouin3,Gavrilyuk1}.
 \\
 For the first time,   deduced from the concept of kinetic energy which introduces an additional variable $\rho$ corresponding to the density, the mass intervenes.
 We add the mass behavior: the mass of any material portion of the medium is conserved along the motions  and is expressed by the   equation  
 \begin{equation*}
 	{\rm Div}(\rho\,\mathcal V)=0
 \end{equation*}
 In some specialized theories (diffusion, aerosols, porous media, etc.) we have to write there is no conservation of mass,  
 \begin{equation*}
 	{\rm Div}(\rho\,\mathcal V)=k,
 \end{equation*}
 where $k$ denotes the source of   mass.
 \\
 The potential energy,  called internal energy $\rho\,\alpha({\boldsymbol G}, s)$, is  chosen in static and is given   by the thermodynamics  which introduces   the specific entropy $s$ which is a   new state variable. The entropy $s$   is not a mechanical   variable and its variation $\delta s$ is   zero \cite{Serrin}. This fact  summarizes  the thermodynamics, and the Kelvin temperature is $\ \displaystyle\theta= {\partial \alpha}/{\partial s}$. The additional variable $s$ being introduced,
 it seems that  an equation is missing. This is not the case, Eq. \eqref{eq entropy} will give   the  evolution of $s$.
 \\
 Equations \eqref{key1} and \eqref{key2} yield $\boldsymbol S^T$ and $\boldsymbol T$ issued from Lagrangian \eqref{Lagrangien2} 
 \begin{equation*}
 	\left | \begin{array}{l}
 		\displaystyle
 		{\rm Source}:\quad \boldsymbol S  = \boldsymbol 0  \\ 
 		{\rm Momentum\ source}:\quad \boldsymbol P =  \rho\,\boldsymbol v\\
 		{\rm  Stress \ tensor}: \quad\boldsymbol \sigma   =   
 		\displaystyle 2\rho \boldsymbol F\,\frac{\partial \alpha}{\partial \boldsymbol G}\, \boldsymbol F^T\\
 		{\rm Total\ energy}:\quad
 		e= \displaystyle \rho\left(\frac{1}{2}\, |\boldsymbol v|^2+ \alpha({\boldsymbol G}, s)\right)\\
 		{\rm Energy\ flux\ vector}:\quad\boldsymbol U = e\,\boldsymbol v- \boldsymbol \sigma \,\boldsymbol v  
 	\end{array}\right.
 \end{equation*}
 For the special case of compressible fluid $\alpha =\alpha(\rho, s)$ and 
$
 \boldsymbol \sigma = -p\,\boldsymbol 1$  where $\displaystyle p= \rho^2\frac{\partial\alpha}{\partial \rho}$  is  the thermodynamic  pressure.\\

 \noindent In the general case, the equations of motion \eqref{delta a1} and \eqref{eq entropy} reduce to
 \begin{equation}
 	{\rm Div} \,\boldsymbol T =\boldsymbol 0\quad{\rm and}\quad \left({\rm Div}\,\boldsymbol T \right) \mathcal V =0 \label{eq bis}
 \end{equation}
 Let's clarify \eqref{eq bis} by decomposing $\boldsymbol T$ into $\boldsymbol T_1$ and $\boldsymbol   T_2$ coming from kinetic energy and potential energy, respectively 
 \begin{equation*}
 	\begin{array}{l}
 		\displaystyle
 		\boldsymbol T_1 = \left[ \begin{matrix}
 			-\frac{1}{2}\rho\,|\boldsymbol v|^2,\ \rho\,\boldsymbol v^T\\ \\
 			-\frac{1}{2}\,|\boldsymbol v|^2  \rho\,\boldsymbol v , \ \   \rho \boldsymbol v\,\boldsymbol v^T
 		\end{matrix}\right]\qquad\qquad \displaystyle
 		\boldsymbol T_2 = \left[ \begin{matrix}
 			\quad-\rho\, \alpha\, ,\qquad \qquad   \boldsymbol 0 ^T\\ \\
 			-\rho\, \alpha\,\boldsymbol v+\boldsymbol \sigma\,\boldsymbol v \,,   -\,\boldsymbol \sigma
 		\end{matrix}\right] 
 	\end{array}
 \end{equation*}
 Then,  $\displaystyle\boldsymbol a = \frac{d\boldsymbol v}{dt}$ denoting the acceleration vector,
 \begin{equation*}
 	\begin{array}{l}
 		\left({\rm Div}\, \displaystyle
 		\boldsymbol T_1\right)^T = \left[ \begin{matrix}
 			-\frac{1}{2}\,|\boldsymbol v|^2{\rm Div} (\rho\,\mathcal V)- \rho\,\boldsymbol v^T\boldsymbol a\\ \\
 			\boldsymbol v\, {\rm Div} (\rho\,\mathcal V) +   \rho\, \boldsymbol a
 		\end{matrix}\right]\  \Longrightarrow\quad\displaystyle \left({\rm Div}\, \displaystyle
 		\boldsymbol T_1\right)^T\,\mathcal V=\frac{1}{2}\,|\boldsymbol v|^2\,{\rm Div} (\rho\,\mathcal V)
 	\end{array}
 \end{equation*}

 \begin{equation*}
 	\begin{array}{l}
 		\displaystyle
 		\left({\rm Div}\, \displaystyle
 		\boldsymbol T_2\right)^T = \left[ \begin{matrix}
 			\quad  -\alpha\, {\rm Div} (\rho\,\mathcal V)-  \rho \displaystyle\frac{d\alpha}{dt}  +{\rm div}(\boldsymbol \sigma\,\boldsymbol v) \\ \\
 			-\left({\rm div}(\boldsymbol \sigma)\right)^T
 		\end{matrix}\right]\qquad  \Longrightarrow \\ \\
 		\left({\rm Div}\, \displaystyle
 		\boldsymbol T_2\right)^T\,\mathcal V=-\alpha\,{\rm Div} (\rho\,\mathcal V)-  \rho \displaystyle\frac{d\alpha}{dt}  +{\rm div}(\boldsymbol \sigma\,\boldsymbol v)-{\rm div}(\boldsymbol \sigma)\,\boldsymbol v\equiv -\alpha\,{\rm Div} (\rho\,\mathcal V)-  \rho \displaystyle\frac{ds}{dt} 
 	\end{array}
 \end{equation*}
 Hence,
 \begin{equation}
 	\left\{\begin{array}{l}
 		\displaystyle
 		\rho\,\frac{d\alpha}{dt}+\rho\, \boldsymbol v^T\boldsymbol a -{\rm div}(\boldsymbol \sigma\,\boldsymbol v)+\frac{1}{2}\,|\boldsymbol v|^2{\rm Div} (\rho\,\mathcal V)=0\label{energy+motion}\\
 		\\
 		\displaystyle \rho\, \boldsymbol a=({\rm div}{\,\boldsymbol \sigma})^T -{\rm Div} (\rho\,\mathcal V)\,\boldsymbol v  
 	\end{array}\right.
 \end{equation}
Equation \eqref{energy+motion}$^1$ is the equation of energy and \eqref{energy+motion}$^2$ is the equation of momentum.  Here, the equation \eqref{eq entropy}  consequence of Eqs. \eqref{energy+motion} yields 
 \begin{equation*}
 	\rho\,\theta \frac{ds}{dt}=\left(\frac{1}{2}\,|\boldsymbol v|^2-\alpha\right) \,{\rm Div} (\rho\,\mathcal V) 
 \end{equation*}
 which is the equation verified by the specific entropy. We have written the equations with terms of mass source  and it is necessary to specify the mechanism by which the mass is introduced. \\ If there is no source of mass,	${\rm Div} (\rho\mathcal V)=0$ and  equations \eqref{energy+motion} reduce  to  
 \begin{equation}
  \rho\,\frac{d\alpha}{dt}+\rho\, \boldsymbol v^T\boldsymbol a -{\rm div}(\boldsymbol \sigma\,\boldsymbol v)=0,\quad \rho\, \boldsymbol a=({\rm div}{\,\boldsymbol \sigma})^T,\quad \frac{ds}{dt}=0\label{keyequations}
 \end{equation}
 \subsection{Non-isolated conservative medium}
Even if this is often neglected, the continuous medium is subject to the influence of exterior and an external action  must be added. The contact forces can be taken into account by a virtual action integrated on the edge $\partial\mathcal W$ and have an effect on the boundary conditions.\\
 If remote actions only occur through source terms  $\boldsymbol S^T =[-\mathcal P, \boldsymbol F^T]$  which have to be known,    equations \eqref{keyequations}   are replaced by 
 \begin{equation}
 	\left\{\begin{array}{l}
 		\displaystyle
 		\rho\,\frac{d\alpha}{dt}+\rho\, \boldsymbol v^T\boldsymbol a -{\rm div}(\boldsymbol \sigma\,\boldsymbol v)= \mathcal P\\
 		\\
 		\displaystyle \rho\, \boldsymbol a=({\rm div}{\boldsymbol \sigma})^T +\boldsymbol F\\
 		\\
 		\displaystyle\rho\,\theta\,\frac{ds}{dt}=\mathcal P-\boldsymbol F^T\boldsymbol v\label{key1equations}
 	\end{array}\right.
 \end{equation}
 In general, the motion is not adiabatic and $(\mathcal P-\boldsymbol F^T\boldsymbol v)/(\rho\,\theta)$ is the source of entropy.\\  
 
\noindent An usual case corresponds to external forces deriving from a potential.  
In this case, it is better to return to Hamilton's principle with a new Lagrangian with the  potential $\Omega(\boldsymbol z)$ per unit of mass
 \begin{equation*}
 	\mathcal L = \rho\left(\frac{1}{2} \boldsymbol v^2- \alpha({\boldsymbol G}, s)-\Omega(\boldsymbol z)\right)
 \end{equation*}
 The Lagrangian is not invariant by spatial translation. Nevertheless, the continuous medium remains conservative. Term $-\rho\,\Omega$ added to the Lagrangian introduces a source term $\boldsymbol S_3$ and a tensor $\boldsymbol T_3$  
 \begin{equation*}
 	\boldsymbol S_3 =-\rho\,\frac{\partial \Omega}{\partial \boldsymbol z}\qquad{\rm and}\qquad\boldsymbol T_3=\left[ \begin{matrix}
 		-\rho\, \Omega   \quad  ,  \  \boldsymbol 0 ^T\\ \\
 		-\rho\, \Omega\,\boldsymbol v,   \ \boldsymbol 0
 	\end{matrix}\right] 
 \end{equation*}
 with 
 \begin{equation*}
 	\mathcal P = \rho\,\frac{\partial \Omega}{\partial t}\qquad{\rm and}\qquad\boldsymbol{F}= -\rho\,{\rm grad}\,\Omega
 \end{equation*}
  We must   add to the first member of Eq. \eqref{delta a1} the term
${Div}\,\boldsymbol{T}_3 -\boldsymbol{S}_3^T  
$.
 As $\left({Div}\,\boldsymbol{T}_3 -\boldsymbol{S}_3^T\right)\,\mathcal V =0$, equation \eqref{eq entropy} is unchanged  and  Eqs. \eqref{key1equations}$^2$ and \eqref{key1equations}$^3$ write
 \begin{equation*}
 	\left\{\begin{array}{l}
 		\displaystyle \rho\, \boldsymbol a=({\rm div}{\,\boldsymbol  \sigma})^T -\rho\, {\rm grad}\,\Omega\\
 		\\
 		\displaystyle\rho\,\theta\,\frac{ds}{dt}=0
 	\end{array}\right.
 \end{equation*}
If we write the total energy $e =1/2\,\rho\,|\boldsymbol v|^2+\rho\, \alpha+\rho\,\Omega$, the equation of energy \eqref{key1equations}$^1$ can be written
 \begin{equation*}
 	\frac{\partial e}{\partial t}+{\rm div}\, \boldsymbol U = \rho\,\frac{\partial\Omega}{\partial t},\quad{\rm where}\quad \boldsymbol U = e\,\boldsymbol v- \boldsymbol \sigma \,\boldsymbol v  
 \end{equation*}
 The motion remains adiabatic. This result is consistent with the thermodynamics: the entropy of a gas in a tank does not change if the tank moves in the gravity field.\\
 
\noindent It is possible to comment on the above results as follows
 
{\bf -}  Or to obtain the total energy\ $e$, one considers the internal energy and an external potential energy which is added to the kinetic energy. Consequently, we have  an energy source $\mathcal P= \rho\,\partial\Omega/\partial t$,   a momentum source $\boldsymbol F =-\rho\  {\rm grad}\,\Omega$ and an additional energy flux vector $\rho\, \Omega\, \boldsymbol v$. 
 
{\bf -}   Or we only consider  the energy in the form $e =1/2\,\rho\,|\boldsymbol v|^2+ \rho\,\alpha$ and we have  an energy source $\mathcal P = \boldsymbol F^T\boldsymbol v$ and a momentum source $\boldsymbol F =-\rho\, {\rm grad}\,\Omega$.
 \\
 In both cases the entropy source, different from the energy source, is zero.
 
 \subsection{Introduction of dissipative terms}
 A supplementary tensor $\boldsymbol T_4$ added to the stress--energy tensor allows to introduce dissipative effects (viscosity, thermal conduction).   
 
 {\bf -} In the case of viscous motion, a viscosity stress tensor $\boldsymbol \sigma_1$ is added to the stress tensor $\boldsymbol \sigma$.
  
 {\bf -}  In the case of thermal conduction, a vector $\boldsymbol U_1$ is added to the energy flux $\boldsymbol{U}$   \cite{Bersani,Ruggeri}.\\ 
 
 \noindent The stress--energy  tensor becomes  
 \begin{equation*}
 	\boldsymbol{T}=\left[\begin{matrix}
 		-e\qquad\qquad\qquad\,,\qquad     \rho \boldsymbol v ^T\\ \\
 		\ \ -e\,\boldsymbol v+(\boldsymbol\sigma+\boldsymbol\sigma_1) \,\boldsymbol v -\boldsymbol U_1 ,\,\  \rho\, \boldsymbol v\,  \boldsymbol v ^T- \boldsymbol \sigma- \boldsymbol \sigma_1   
 	\end{matrix}\right] 
 \end{equation*}
 $\displaystyle
 {\rm with}\quad\boldsymbol{T}_4=\left[\begin{matrix}
 		\quad 0\quad\qquad\,\ ,  \quad  \boldsymbol 0^T\\ 
 		\ \   \boldsymbol\sigma_1\boldsymbol v -\boldsymbol U_1 ,  -\, \boldsymbol \sigma_1   
 	\end{matrix}\right]
$.\\

\noindent If $\boldsymbol U_1$ is zero, note that $\mathcal V$ remains an eigenvector of the stress--energy tensor.  We deduce 
 \begin{equation*}
 	({\rm Div}\, T_4)\,\mathcal V=\Psi-{\rm div}\, \boldsymbol U_1\quad{\rm with}\quad \Psi = {\rm div} (\boldsymbol\sigma_1\boldsymbol v)-{\rm div} (\boldsymbol\sigma_1)\boldsymbol v\equiv {\rm Tr}\left(\boldsymbol\sigma_1\frac{\partial\boldsymbol v}{\partial\boldsymbol x}\right)
 \end{equation*}
 Scalar $\Psi$ is  the dissipation function or the power dissipated by the viscosity  per unit volume. In the case of isolated media, the equations of motion are 
 \begin{equation}
 	\left\{\begin{array}{l}
 		\displaystyle
 		\frac{\partial e}{\partial t}  +{\rm div}(e\,\boldsymbol v-\boldsymbol \sigma\,\boldsymbol v)= {\rm div} (\boldsymbol\sigma_1\boldsymbol v)-{\rm div}\,\boldsymbol{U}_1\\
 		\\
 		\displaystyle \rho\, \boldsymbol a={\rm div} (\boldsymbol \sigma^T +\boldsymbol \sigma^T_1) \\
 		\\
 		\displaystyle\rho\,\theta\,\frac{ds}{dt}=\Psi-{\rm div}\,\boldsymbol{U}_1
 	\end{array}\right.\label{Neweq}
 \end{equation}
 to which it may be necessary to add source terms for non-isolated media.
 \section{Behavior laws for dissipative forces}
 
 The equations \eqref{Neweq} only constitute a general framework. It is necessary to present   behavior laws characterizing the medium and allowing to obtain $\boldsymbol \sigma_1$ and $\boldsymbol U_1$ as function of the motion. The second law of thermodynamics is the only way to limit the possible variety of the behaviors \cite{Truesdell}.\\
 The entropy of  isolated systems can only  increase (or remain constant). Another restriction is that the entropy increase occurs at each point, i.e., 
 \begin{equation*}
 	\forall\, \boldsymbol  z \in \mathcal W,\quad \frac{ds}{dt} \geq 0
 \end{equation*}
 If we also require that each term constituting the entropy variation remains positive  at each point and at each time, we obtain 
 \begin{equation*}
 	\Psi \geq 0 \qquad{\rm and}\qquad {\rm div}\,\boldsymbol U_1\leq 0 
 \end{equation*}
 It remains to classify dissipative media, according to the behavior they may have with respect to spatial and material invariance. We only describe  the simplest of dissipative media, i.e., the viscous fluids.
 
 $1^\circ)$ $\boldsymbol\sigma_1$ is an isotropic function of the deformation rate ${\boldsymbol D}$ so that 
 \begin{equation*}
 	{\boldsymbol D} =\frac{1}{2} \left(\frac{\partial {\boldsymbol v}}{\partial x}+\left(\frac{\partial {\boldsymbol v}}{\partial x}   \right)^T\right)
 \end{equation*}  
 In the linearised case,
 \begin{equation*}
 	\boldsymbol \sigma = \lambda\,{\rm Tr}(	{\boldsymbol D})+ 2\,\mu 	{\boldsymbol D} ,
 \end{equation*}  
 and the dissipation function is
 \begin{equation*}
 	\Psi = \lambda\,({\rm Tr}\,	{\boldsymbol D})^2+ 2\,\mu {\rm Tr}(	{\boldsymbol D}^2)
 \end{equation*}  
The condition $\Psi\geq 0$ reduces to $3\,\lambda+2\,\mu \geq 0$ and $\mu \geq 0$.\\
 Terms $\lambda$ and $\mu$  are the dynamic viscosity coefficients, whose values can be found for the different gases and liquids in the tables of physical constants.
 For the sake of simplicity, the Stokes hypothesis $3\,\lambda+2\,\mu = 0$ is sometimes used, which   assumes that there is no energy dissipation if a gas is expanding spherically.\\
 For incompressible fluids,  ${\rm Tr}\, ({\boldsymbol D}) =0$; then $\boldsymbol{\sigma}_1 = 2\mu\, {\boldsymbol D}$,\ $\Psi = 2\mu\, {\rm Tr} ({\boldsymbol D}^2)$ and only the $\mu$ coefficient is involved. 
 
 $2^\circ)$ The vector $\boldsymbol{U}_1$ represents the energy flux coming from a non-mechanical but thermal action. Since the heat flux goes from hot to cold, for reasons of isotropy we take $\boldsymbol{U}_1$  proportional to the temperature gradient  
 \begin{equation*}
 	\boldsymbol{U}_1 = -k \  {\rm grad}\, \theta,
 \end{equation*}
 where $k >0$ is the heat conduction coefficient. The entropy equation writes:
 \begin{equation*}
 	\rho\,\theta\,\frac{ds}{dt}=\Psi+{\rm div}\,(k \ {\rm grad}\, \theta)
 \end{equation*}
 \section{Shock waves}
 For conservative media, the principle of virtual action is reduced to Hamilton's principle which is  valid for the case of shocks  \cite{Gouin3,Gavrilyuk1}.  
 A shock wave is a moving surface, thus a three-dimensional   manifold $\Sigma$ in the space--time $\mathcal W$, on which the stress--energy tensor is discontinuous.
 We can always write the principle of virtual action. The integration by parts expressed in Section 3  relation \eqref{IPP} must be modified as follows
 \begin{equation*}
 	\delta a =\int_{\mathcal W}\left[\boldsymbol S ^T-   {\rm Div}\left(\boldsymbol T\right)  \right] \boldsymbol \zeta\,dw+\int_{\partial\mathcal W}\boldsymbol N^T\boldsymbol T\,\boldsymbol \zeta\, d\tau+\int_{\Sigma}\boldsymbol N^T\left[\boldsymbol T_2-\boldsymbol T_1\right]\,\boldsymbol \zeta\, d\tau
 \end{equation*}
 In the integral on $\Sigma$, $\boldsymbol T_1$ and  $\boldsymbol T_2$ are the values of  $\boldsymbol T$ on each side of $\Sigma$, $\boldsymbol N^T$ is the covector, annulator of vectors in the hyperplane tangent to $\Sigma$.
 \\
 We have $\boldsymbol N^T= [-D_n, \boldsymbol n^T]$.
 where $\boldsymbol{n}$ is the downstream unit normal vector to the shock wave  and $D_n$ is normal velocity of the shock wave.\\
 We deduce from the principle of virtual action 
 
 $1^\circ)$ Outside from $\Sigma$, the equations of motion \eqref{delta a1} are unchanged.
 
 $2^\circ)$ On $\Sigma$ we have the shock condition 
 $\boldsymbol N^T\left[\boldsymbol T_2-\boldsymbol T_1\right]=\boldsymbol 0^T$.
 \\ 
We note that on $\Sigma$ we automatically obtain the conservation of momentum and the conservation of energy which is classically postulated  \cite{Gouin3}.

 \section{Conclusion and comments}
The principle of virtual work is convenient for studying the statics of continuous media.
The principle of virtual action is suitable for dynamics.\\ In this paper, we present the classical case of fluid mechanics and elasticity. This presentation can be extended to more complex media as it is done by the principle of virtual work in the case of second gradient models, capillary fluids, metamaterials, etc \cite{Del Vescovo,Auffray,dellIsola}. The principle of virtual action allows us to directly obtain the equations of motion and energy and makes explicit the possibility of shock waves for dispersive media  \cite{Gavrilyuk2}.\\
 
 \noindent {{\textbf{Acknowledgments :}\\
 		{This article is dedicated to the memory of Professor Salvatore Rionero whose research has greatly contributed  to the foundation of theoretical models. \\ 
 			This work was partially supported by National Group of Mathematical Physics
 			GNFM-INdAM (Italy).}


\begin{thebibliography}{99}
 			
 					\bibitem{Serrin} Serrin, J.: Mathematical principles of classical fluid mechanics. In: Flügge, S. (Ed.) Encyclopedia of
 			physics VIII/1, p.p. 125--263, Springer, Berlin (1960).
 			
 			\bibitem{Casal} Casal, P.: Principes variationnels en fluide compressible et en magn\'etodynamique des fluides, Journal de Mécanique, 
 			{\bf 5}, 149--161 (1966).
 			
 			\bibitem{Marsden}	Marsden, J.E., Hughes, T.J.R.:  Mathematical Foundations of Elasticity, Dover Publications, New York (1983).
 			
 				\bibitem{Gouin1} Gouin, H.: The d’Alembert–Lagrange principle for gradient theories and boundary conditions, in:  Ruggeri,T. and Sammartino, M. (Eds.), Asymptotic
 			Methods in Nonlinear Wave Phenomena, World Scientific, Singapore, p.p. 79–95 (2007).
 			
 				\bibitem{Germain}	Germain, P.: The method of virtual power in the mechanics of continuous media, I: Second gradient theory. Mathematics and Mechanics of Complex Systems, {\bf 8}, 153–190 (2020), translation from  \textit{ La méthode des puissances virtuelles en mécanique des milieux continus, premi\`ere partie: théorie du second gradient, Journal de Mécanique,}  {\bf 12},  235--274 (1973).
 				
 					
 				\bibitem{Gouin3} Gouin, H.: Rankine--Hugoniot conditions obtained by using the
 				space--time Hamilton action. Ricerche di Matematica, {\bf 70}, 115--129 (2021).
 			
 			\bibitem{Gavrilyuk1} 	Gavrilyuk, S., Gouin, H.: New form of governing equations of fluids arising from Hamilton’s principle,
 			Int. J. Eng. Sci., {\bf 37}, 1495--1520 (1999) 
 			
 				\bibitem{Gavrilyuk2} 	Gavrilyuk, S., Gouin, H.:	Rankine–Hugoniot conditions for fluids whose energy	depends on space and time derivatives of density, Wave Motion, {\bf 98}, 102620  (2020).
 			
 			\bibitem{Souriau} Souriau, J.-M., G\'eom\'etrie et Relativit\'e, Editions Hermann, Paris (1964).
 			
 			
 			\bibitem{Lamb}Lamb, H.: Hydrodynamics, Dover Publ., New York (1972).
 			
 			\bibitem{Dafermos}	Dafermos, C.: Conservation Laws in Continuum Physics, 2nd edn., Springer, Berlin (2005).
 			
 				\bibitem{Capobianco}	Capobianco, G., Winandy, T., Eugster, S.: The principle of virtual work and Hamilton’s principle on Galilean manifolds, Journal of Geometric Mechanics, {\bf 13},  167--193 (2021).
 			
 				\bibitem{Bersani}	Bersani, A.M., Caressa, P.: Lagrangian descriptions of dissipative systems: a review,  
 			Math. Mech. Solids,
 			{\bf 26}, 785--803 (2021).
 			
 			\bibitem{Ruggeri} Ruggeri, T., Sugiyama, M.: Classical and Relativistic Rational Extended Thermodynamics of Gases, Springer, Switzerland (2021).
 			
 			\bibitem{Truesdell}  Truesdell, C.: Rational Thermodynamics, McGraw Hill, New-York (1969).
 			
	
		
	
			
 	
		\bibitem{Del Vescovo}	del Vescovo, D., Giorgio, I.: Dynamic problems for metamaterials: review of existing models and ideas for further research, Int. J. Eng. Sci., {\bf 80},  153--172 (2014).
		
			\bibitem{Auffray}	Auffray, N., dell’Isola, F., Eremeyev, V.A., Madeo, A., Rosi, G.: Analytical continuum mechanics \textit{à la} Hamilton-Piola least action principle for second gradient continua and capillary fluids, Math. Mech. Solids, {\bf 20}, 375--417 (2015).
 		
 			\bibitem{dellIsola}	dell'Isola, F., Seppecher, P., Placidi, L., Barchiesi, E., Misra, A.:  Least Action and Virtual Work Principles for the Formulation of Generalized Continuum Models, In:  {Discrete and Continuum Models for Complex Metamaterials}, p.p. 327--394, Cambridge University Press, Cambridge (2020).
		 
 	
% 			
 			
 			
 			
 			
 			
 			
 			
 			
 			
 			
 			
 			
 			
 			
 			
 			
 			
 			
 			
 			
 			
 		\end{thebibliography}
 	\end{document}